\documentclass[aps,showpacs]{revtex4}
\usepackage{graphicx,psfrag}
\def\no{\noindent}
\def\bc{\begin{center}}
\def\ec{\end{center}}

\def\beq{\begin{equation}}
\def\eeq{\end{equation}}

\begin{document}

\title{
Weak-localization approach to a 2D electron gas with a spectral node
}

\author{K. Ziegler and A. Sinner}
\affiliation{Institut f\"ur Physik, Universit\"at Augsburg\\
D-86135 Augsburg, Germany\\
}
\date{\today}

\begin{abstract}
We study a weakly disordered 2D electron gas with two bands and a spectral node within the weak-localization approach and 
compare its results with those of Gaussian fluctuations around the self-consistent Born approximation. 
The appearance of diffusive modes depends on the type of disorder. In particular, we find
for a random gap a diffusive mode only from ladder contributions, whereas for a random scalar potential
the diffusive mode is created by ladder and by maximally crossed contributions. The ladder (maximally crossed) 
contributions correspond to fermionic (bosonic) Gaussian fluctuations. 
We calculate the conductivity corrections from the density--density Kubo formula and find a good
agreement with the experimentally observed V-shape conductivity of graphene.
\end{abstract}
\pacs{05.60.Gg, 66.30.Fq, 05.40.-a}

\maketitle

\section{Introduction}

The weak-localization approach (WLA) has been a very popular tool to estimate whether electronic 
states in a weakly disordered
system tend to localize or to delocalize on large scales. Moreover, it enables us to calculate the magnetoresistance
in the presence of a weak magnetic field and weak scattering. A central result of the WLA is that on large scales there
might be diffusion due to one or more massless modes. This has been studied in great detail for conventional metals
\cite{gorkov79,hikami80,altshuler} and more recently for graphene \cite{andoetal,khveshchenko06}
and for the surface of topological insulators \cite{tkachov11,schmeltzer13}, using a one-band projection for the two-band system. T
he existence  of a diffusive mode, which is a necessary 
(but not a sufficient) condition for metallic behavior, has been debated for the one-band projected graphene model.
It was found that either a single diffusive channel exists \cite{andoetal,tkachov11,schmeltzer13} or no diffusion
\cite{khveshchenko06} in the presence of generic disorder. On the other hand, the weak-scattering approach (WSA), 
where transport properties are studied within the expansion in powers of $\eta/E_b$ ($\eta$ is the scattering rate
and $E_b$ is the bandwidth) \cite{ziegler97}, a non-Abelian chiral symmetry was identified, describing diffusion in two-band systems
due to spontaneous symmetry breaking \cite{ziegler12b}. This is also the origin of a massless fermion mode found for 
2D Dirac fermions with a random gap in Ref. \cite{ziegler98}. 
In the WSA disorder fluctuations of the two-band model are approximated by Gaussian fluctuations around a saddle-point 
of the original model, expressed in terms of a functional integral \cite{ziegler97}. The saddle point is equivalent to the
self-consistent Born approximation (SCBA) of the one-particle Green's function, while the Gaussian fluctuations 
are equivalent to the WLA. The latter consists of one-particle
and two-particle diagrams which are partially summed up in terms of geometric series (cf. Sect. \ref{sect:wla}). 
Within the WSA it is also possible to analyze the fluctuations with respect to the non-Abelian chiral symmetry. 
The projection onto these fluctuations generates a nonlinear field which allows us to go beyond the Gaussian 
approximation within the expansion in powers of $\eta/E_b$. This idea is analogous to the nonlinear sigma model,
derived originally for one-band Hamiltonians by Sch\"afer and Wegner \cite{wegner80}. The difference between the 
one-band and the two-band Hamiltonians is that the former can be formulated either by a symmetric replica space or a 
supersymmetric fermion-boson \cite{efetov97}, whereas the latter can be expressed in terms of a non-symmetric 
fermion-boson theory \cite{ziegler98}. Therefore, in the derivation of a nonlinear sigma model it is crucial to take the 
two-band structure into account. A projection onto a single band would destroy the relevant symmetries
of the system. In more physical terms, the two-band structure is essential for supporting diffusion in a two-dimensional
system, since it allows for Klein tunneling. The latter enables a particle in a potential barrier to transmute to
a hole, for which the potential barrier is not an obstacle. Our aim is to establish a direct connection between the
WLA and the Gaussian fluctuations around the saddle point for 2D Dirac fermions with a random gap, and
to provide a general discussion about the existence of diffusive modes due to ladder and maximally crossed
contributions in two-band systems. Finally, these results will be employed to calculate the conductivity corrections,
and the resulting conductivities will be compared with experimental measurements in graphene. The results can also be applied 
to other 2D two-band systems such as the surface of topological insulators \cite{review_ti}.

The paper is organized as follows. In Sect. \ref{sect:model} we introduce a general description for the two-band Hamiltonian
and various types of random scattering. The main ideas of the WLA are discussed in Sect. \ref{sect:wla}, which includes the 
self-consistent Born approximation for the average one-particles Green's function, the ladder and the maximally crossed 
contribution of the average two-particle Green's function. In Sect. \ref{sect:special} we study the long-range behavior
of the average two-particle Green's function for a one-band Hamiltonian (Sect. \ref{sect:scalar}) and for the two-band
Hamiltonian (Sect. \ref{sect:spinor_gf}). These results are used to calculate the conductivity (Sect. \ref{sect:correction}).
And finally, in Sect. \ref{sect:discussion} we discuss the connection of the WLA with the WSA, the robustness of the diffusion
pole structure with respect to a one-band projection of the two-band Hamiltonian and the symmetry properties of the 
inter-node scattering.

\section{Model: Hamiltonians, Green's functions and symmetries}
\label{sect:model}

Quasiparticles in a system with two bands are described by a spinor wavefunction.
The corresponding Hamiltonian can be expanded in terms of Pauli matrices $\sigma_{0,1,2,3}$.
Here we will consider either a gapless Hamiltonian
\beq
H_0=h_1\sigma_1+h_2\sigma_2
\eeq
or a gapped Hamiltonian
\beq
H_m=h_1\sigma_1+h_2\sigma_2+m\sigma_3
\ .
\label{gap_ham}
\eeq
The gapless Hamiltonian changes its sign under a chiral transformation 
\beq
\sigma_3 H_0\sigma_3=-H_0
\ ,
\label{ham0}
\eeq
which implies the continuous Abelian chiral symmetry
\[
e^{\alpha\sigma_3}H_0 e^{\alpha\sigma_3}=H_0
\ .
\]
The situation is more subtle for $H_m$ because its transformation properties depends on the properties of $h_{1,2}$.
We distinguish here two cases, namely $h_j^T=-h_j$ (Dirac fermions, $^T$ is the transposition, acting on real space), where
\beq
\sigma_1 H_m^T\sigma_1=-H_m
\label{ham1}
\eeq
and $h_j^T=h_j$ (e.g., bilayer graphene), where
\beq
\sigma_2 H_m^T\sigma_2=-H_m
\ .
\label{ham2}
\eeq
The transformation properties of the Hamiltonians imply a relation between the wavefunctions in the upper and in the lower
band. In particular, Eq. (\ref{ham0}) implies that $\Psi_{-E}=\sigma_3\Psi_E$, Eq. (\ref{ham1}) implies
that $\Psi_{-E}=\sigma_1\Psi^*_E$ and Eq. (\ref{ham2}) that $\Psi_{-E}=\sigma_2\Psi^*_E$.

Disorder is included by an additional random term $diag(v_1,v_2)$ in the Hamiltonians $H_0$ and $H_m$. 
In the following we will consider the case of $v_1=v_2$ (scalar potential), $v_1=-v_2$ (random gap) and independent
random diagonal elements $v_1$, $v_2$, 
assuming that the matrix elements $v_{1,2}$ have mean zero and variance $g$.

\section{Dyson and Bethe-Salpeter equation}
\label{sect:wla}

\subsection{Dyson equation}
\label{sect:dyson}

Starting point of the WLA is that the Green's function
$G_0(\mu-i\delta)=(H_0-\mu+i\delta)^{-1}$ 
of the Hamiltonian $H_0$ is perturbed by $V$ and creates the Green's function $G=(G_0^{-1}+V)^{-1}$. 
The latter relation is equivalent to the matrix identity 
$
G=G_0-G_0VG
$
which connects the unperturbed Green's function $G_0$ with the perturbed Green's function $G$. 
This equation can be iterated to give a geometric series
\beq
G=G_0-G_0VG=G_0-G_0VG_0+G_0VG_0VG = ... = G_0\sum_{l\ge 0}(-VG_0)^l
\ .
\label{dyson2}
\eeq
Now we assume that $V$ is a random quantity with mean $\langle V\rangle=0$ such that the averaged Dyson equation
becomes
\beq
\langle G\rangle =G_0+G_0\langle VG_0VG\rangle
\ .
\eeq
When we assume that the correlation between the Green's function and $V$ is weak, the factorization of the average 
$\langle VG_0VG\rangle\approx \langle VG_0V\rangle\langle G\rangle$ is possible, which creates a linear equation for 
$\langle G\rangle$:
\beq
\langle G\rangle \approx G_0+G_0\langle VG_0V\rangle\langle G\rangle
\ ,
\label{dyson3}
\eeq
whose solution reads $\langle G\rangle\approx 
(G_0^{-1}-\langle VG_0V\rangle)^{-1}$. 
This result is known as the Born approximation with the self-energy $ \langle VG_0V\rangle$. 
The self-consistent Born approximation (SCBA) is provided by the replacement 
\beq
\langle V_iG_{0,ij} V_j\rangle \to \langle V_i V_j\rangle \langle G_{ij}\rangle 
=:\Sigma_{ij}  
\label{scba}
\eeq
on the right-hand side of Eq. (\ref{dyson3}):
\beq
\langle G\rangle\approx {\bar G}=(G_0^{-1}-\Sigma)^{-1} 
\ .
\eeq

\subsection{Bethe-Salpeter equation}

The two-particle Green's function $G^+G^-$, created from the two one-particle Green's functions $G^\pm$
(e.g., the adavanced and the retarded Green's function),
reads with the help of the corresponding Dyson equations (\ref{dyson2})  (using the summation convention in this Section)
\beq
\langle G^+_{ij}G^-_{kl}\rangle =\langle T^{-1}_{ik,mn}\rangle G^+_{0,mj}G^-_{0,nl} \ \ \ {\rm with}\ \ 
T_{ik,mn}= \delta_{im}\delta_{kn} - G^+_{0,im'}V_{m'm} G^-_{0,kn'}V_{n'n}
\ .
\label{bse01}
\eeq
On the left-hand side is the two-particle Green's function, while the
right-hand side depends only on products of $G_0$. 
This equation is known as the Bethe-Salpeter equation.
Now we can perform the average with respect to the random scatterers $V_{jj'}$, assuming that
this is a Gaussian variable with zero mean.
Here it is convenient to define $\gamma^\pm_{im}=G^\pm_{0,im'}V_{m'm}$ such that
\beq
T_{ik,mn}= \delta_{im}\delta_{kn} -\gamma^+_{im}\gamma^-_{kn}
\ .
\eeq
The expansion of $T^{-1}$ leads to a geometric series in $\gamma^+_{m_1m_2}\gamma^-_{n_1n_2}$.
Averaging this series  with respect to a Gaussian distribution can be achieved
by what is known as Wick's theorem:
the average of the product of random variable $\langle V_m V_n\cdots \rangle$ is expressed as a sum over all possible
products of pairs $\langle V_m V_n\rangle \cdots$. This series includes a ladder contribution 
and a maximally crossed contribution \cite{langer} (cf. Fig. \ref{fig:diagrams}) as special cases.
We also include the contribution from the iterated Dyson equation of Sect. \ref{sect:dyson} in terms
of the SCBA and obtain eventually
\beq
\langle G^+_{ij}G^-_{kl}\rangle \approx 
[\langle T^{-1}_{ik,mn}\rangle_L+ \langle T^{-1}_{ik,mn}\rangle_M-\delta_{im}\delta_{kn} -t_{ik,mn}]
{\bar G}^+_{mj}{\bar G}^-_{nl}
\ ,
\label{bse02}
\eeq
where the last two terms on the right-hand side are introduced to avoid overcounting in the geometric series.
\begin{figure}[t]
\includegraphics[width=5cm]{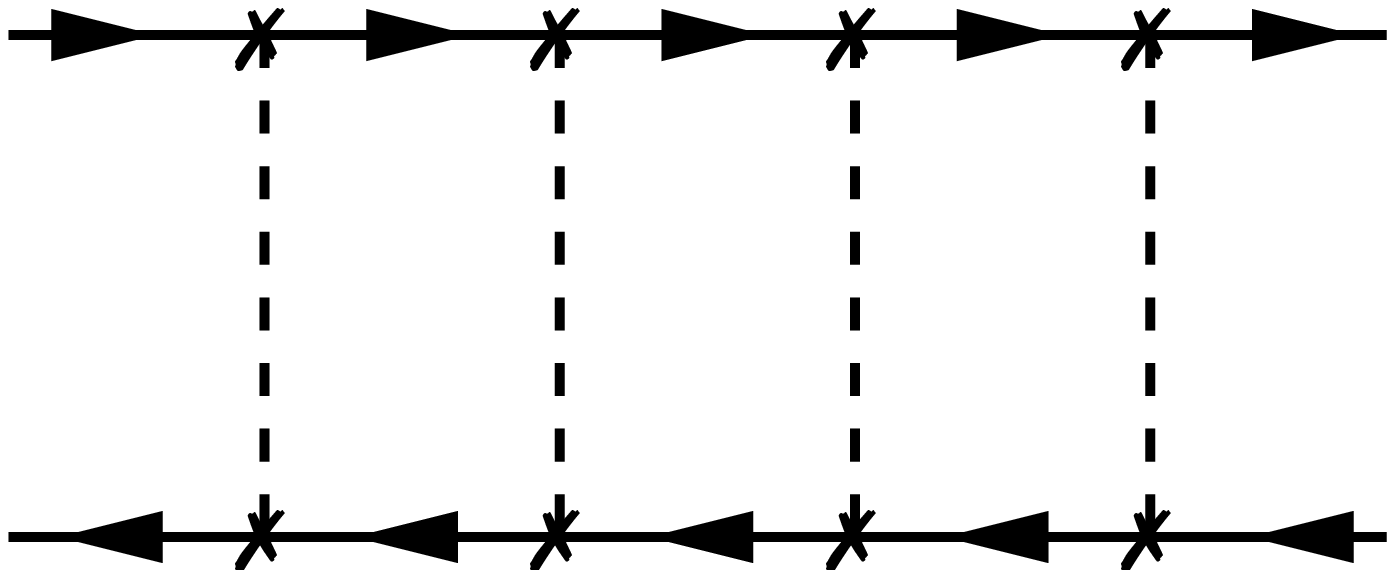} 
\includegraphics[width=5cm]{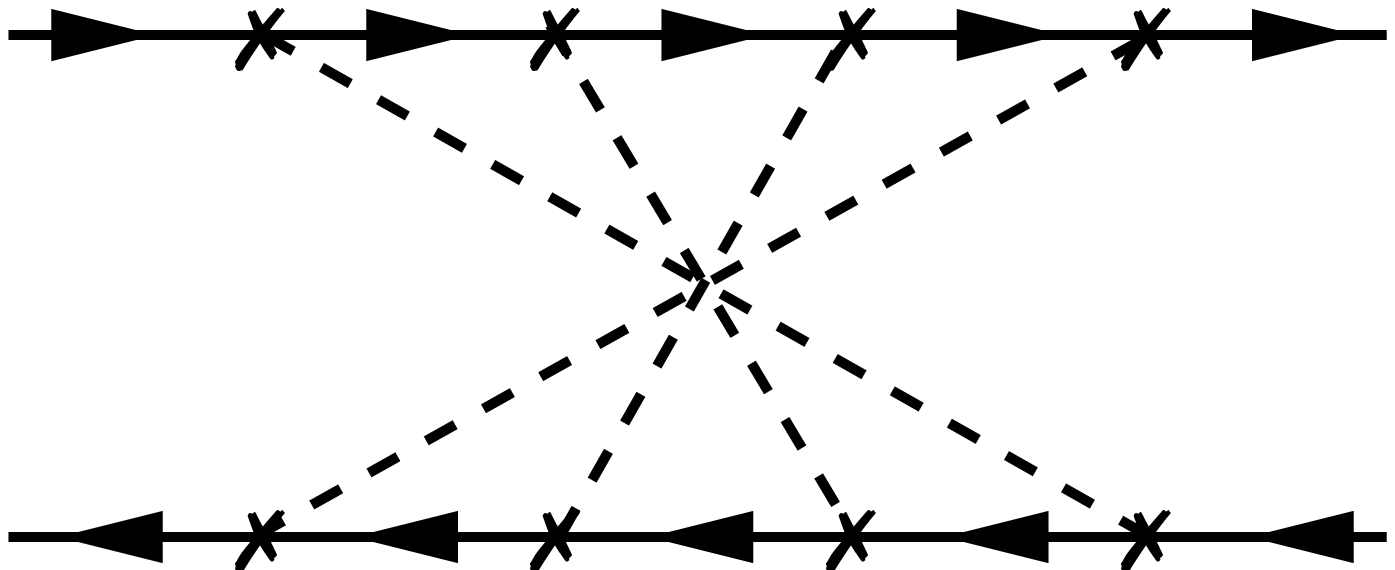} 
\label{fig:diagrams}
\caption{Diagrammatic representation of the fourth order terms for the ladder contribution of $({\bf 1}-t)^{-1}$ and maximally crossed 
contributions of $({\bf 1}-\tau)^{-1}$, respectively.}
\label{fig:diagrams}
\end{figure}
Beginning with the ladder contribution, we obtain
\beq
\langle T^{-1}_{ik,mn}\rangle_L
=({\bf 1}-t)^{-1}_{ik,mn} \ \ {\rm with}\ \  t_{m_1n_1,m_2n_2}
=\langle {\bar G}^+_{m_1m_1'}V_{m_1'm_2}{\bar G}^-_{n_1n_1'}V_{n_1'n_2}\rangle
\ .
\label{series2}
\eeq
Then the maximally crossed contribution is created by re-arranging the order of factors in the geometric series
which after averaging results in
\beq
\langle T^{-1}_{ik,mn}\rangle_M
= ({\bf 1}-\tau)^{-1}_{in,mk} \ \ {\rm with}\ \   \tau_{m_1m_1',m_2m_2'}
=\langle {\bar G}^+_{m_1n_1}V_{n_1m_2} {\bar G}^-_{m_2'n_1'}V_{n_1'm_1'}\rangle
\ .
\label{series4}
\eeq
Thus, switching from the ladder contribution to the maximally crossed contribution is achieved simply through changing 
$\gamma^-$ by the transposition $\gamma^-_{m_1'm_2'}\to \gamma^-_{m_2'm_1'}$.

In the following we will discuss the two-particle Green's functions of (\ref{bse02})--(\ref{series4}) 
and the SCBA for the two-band Hamiltonians of Sect. \ref{sect:model} 
with an additional random term $diag(v_1,v_2)$, which is either a random scalar potential, a random gap or independent random 
diagonal elements.
For simplicity it is assumed that the random terms are spatially uncorrelated.

\section{Special cases of the two-particle Green's function}
\label{sect:special}

The general expressions in (\ref{series2}) and in (\ref{series4}) shall now be applied to specific cases.
For a diagonal random matrix $V_{mm'}=V_m\delta_{m,m'}$ we obtain
\beq
t_{m_1n_1,m_2n_2}
= {\bar G}^+_{m_1m_2}{\bar G}^-_{n_1n_2}\langle V_{m_2}V_{n_2}\rangle , \ \ \ 
\tau_{m_1n_1,m_2n_2}
={\bar G}^+_{m_1m_2} {\bar G}^-_{n_2n_1}\langle V_{m_2}V_{n_1}\rangle
\ .
\label{t-matrix}
\eeq

\subsection{Scalar Green's function}
\label{sect:scalar}

Before we start to discuss the two-band Hamiltonians, the simpler one-band Hamiltonian
with Fourier components $h(k)$ is used to explain briefly the main ideas of the WLA. 
For this case we consider the following coordinates in real space: $i=k=r$, $m=n=r'$.
Then the ladder and the maximally crossed contributions read for uncorrelated disorder 
$\langle V_r V_{r'}\rangle =g\delta_{r,r'}$
\beq
t_{r_1r_2,r_3r_4}=g{\bar G}^+_{r_1r_3}{\bar G}^-_{r_2r_3}\delta_{r_3,r_4}  , \ \ \  
\tau_{r_1r_2,r_3r_4}=g{\bar G}^+_{r_1r_3}{\bar G}^-_{r_4r_3}\delta_{r_2,r_3}
\ ,
\eeq
and in the geometric series $\sum_{l\ge 0}t^l_{rr,r'r'}$ ($\sum_{l\ge 0}\tau^l_{rr',r'r}$)
only
\beq
t_{r_2 r_3}:=t_{r_2r_2,r_3r_3}=g{\bar G}^+_{r_2r_3}{\bar G}^-_{r_2r_3} ,\ \ \ 
\tau_{r_2 r_3}:=\tau_{r_2r_3,r_3r_2}=g{\bar G}^+_{r_2r_3}{\bar G}^-_{r_2r_3}
\label{t-matrix2}
\eeq
contributes. Thus Eq. (\ref{bse02}) reads
\beq
\langle G^+_{rr'}G^-_{rr'}\rangle \approx 
\sum_{r''}[({\bf 1}-t)^{-1}_{rr''}+({\bf 1}-\tau)^{-1}_{rr''}-\delta_{rr''} -t_{rr''}]{\bar G}^+_{r''r'}{\bar G}^-_{r''r'}
\ .
\eeq
Moreover, with (\ref{t-matrix2}) the extra factor
\beq
{\bar G}^+_{r''r'}{\bar G}^-_{r''r'}=\frac{1}{g}t_{r''r'}=\frac{1}{g}\tau_{r''r'}
\ ,
\eeq
can be replaced by $t$ and $\tau$, respectively, to provide
\beq
\langle G^+_{rr'}G^-_{rr'}\rangle \approx 
\frac{1}{g}({\bf 1}-t)^{-1}_{rr'}+\frac{1}{g}({\bf 1}-\tau)^{-1}_{rr'}-\frac{1}{g}(t^2_{rr'}+t_{rr'}+2\delta_{rr'}) 
\ .
\eeq
The Green's function is symmetric in the absence of a magnetic field, such that the Fourier components
of the Hamiltonian $h(k)$ satisfy the relation $h(-k)=h(k)$ . Then there is no difference between
ladder and maximally crossed contributions.
We consider $q\sim 0$ for the long-range behavior and obtain from the expansion in powers of the momentum $q$
\beq
t_q\sim g\sum_r {\bar G}^+_{r0}{\bar G}^-_{0r}-q^2\frac{g}{2}\sum_r r_\mu^2 {\bar G}^+_{r0}{\bar G}^-_{0r}
\eeq
and use the SCBA with the self-energy $\Sigma^\pm$:
${\bar G}^+=G_0(z)$, ${\bar G}^-=G_0(z)^*$ with $z=\mu-i\delta+\Sigma^+$. This gives $t_0\sim  1-2i\delta/(z-z^*)$,
such that $1/(1-t_q)$ becomes the diffusion propagator
\beq
\frac{1}{1-t_q}\sim \frac{1}{2i\delta/(z-z^*)+q^2(g/2)\sum_r r_\mu^2 {\bar G}^+_{r0}{\bar G}^-_{0r}}
\ .
\eeq
This result is remarkable because it implies that $1/(1-t_q)$ diverges like $q^{-2}$ for $q\sim0$ and $\delta\sim 0$, 
reflecting the well-known massless two-particle mode for diffusion \cite{gorkov79,hikami80,altshuler}.

\subsection{Spinor Green's function}
\label{sect:spinor_gf}

For the spinor Hamiltonian $H_m$ of Eq. (\ref{gap_ham}) we introduce the coordinate $r$ and the Pauli matrix
index $a=1,2$ and adopt the same procedure as for the scalar case. With the correlation 
$\langle V_{r,a} V_{r',a'}\rangle =g_{aa'}\delta_{r,r'}$ and with ${\bar G}^+=G_0(z)$, ${\bar G}^-=G_0(z)^*$ for
$z=\mu-i\delta-i\eta$, where $\mu$ is the renormalized Fermi energy (i.e., the bare Fermi energy which is 
shifted by the real part of the self-energy $\Sigma^+$) and $\eta$ is the scattering rate (i.e., the 
imaginary part of the self-energy). Now we use the Fourier representation of $H_m$ and get with $h^2=h_1^2+h_2^2$
for the Green's function
\beq
G_{0,k}(z)
=\frac{-1}{z^2-m^2-h^2}(z\sigma_0+h_1\sigma_1+h_2\sigma_2+m\sigma_3)
\ ,
\label{1pgf0}
\eeq
and
\beq
t_{q;ab,cd}=g_{cd}\int_k G_{0;k,ac} G_{0;q+k,bd}^*
, \ \ \ 
\tau_{q;ab,cd}=g_{cb}\int_k G_{0;k,ac}G_{0;q+k,db}^*
\label{matrices1}
\ .
\eeq

For random scalar potential $v\sigma_0$ (random gap $v\sigma_3$) the prefactors read
$g_{11}=g_{22}=g_{12}=g_{21}\equiv g$ ($g_{11}=g_{22}=-g_{12}=-g_{21}\equiv g$), and 
for independent random diagonal elements $g_{11}=g_{22}\equiv g$, $g_{12}=g_{21}=0$.
Then we get from Eq. (\ref{matrices1}) for $q=0$ the matrices
\beq
t_0=\pmatrix{
\alpha_1 & 0 & 0 & \beta \cr
0 & s\alpha_2 & 0 & 0 \cr
0 & 0 &  s\alpha_3 & 0 \cr
\beta & 0 & 0 & \alpha_4 \cr
}, \ \ \ 
\tau_0=
\pmatrix{
\alpha_1 & 0 & 0 & 0 \cr
0 & s\alpha_2 & \beta & 0 \cr
0 & \beta & s\alpha_3 & 0 \cr
0 & 0 & 0 & \alpha_4 \cr
}
\ ,
\label{2pgf_mb}
\eeq
where $s=-1$ for a random gap,  $s=1$ for random scalar potential, $s=0$ for independent random diagonal elements.
For Dirac fermions we have $h_j=k_j$ and the matrix elements are the following expressions:
\[
\alpha_1=gI(z+m)(z^*+m) ,\  \alpha_2=gI(z+m)(z^*-m) ,\  \alpha_3=gI(z-m)(z^*+m) , \  \alpha_4=gI(z-m)(z^*-m) 
\]
\beq
I=\int_k\frac{1}{|z^2-m^2-k^2|^2} , \ \ \beta=g\int_k\frac{k^2}{|z^2-m^2-k^2|^2}
\ .
\eeq
Using the SCBA we obtain the integral
\beq
\int_k\frac{|z|^2+m^2+k^2}{|z^2-m^2-k^2|^2}=\frac{1}{g}-\frac{\delta}{g\eta}
\ ,
\eeq
which implies $\beta\sim 1-gI(|z|^2+m^2)$. 
Thus, besides two independent diagonal matrix elements,
we get two eigenvalues for the non-diagonal $2\times2$ submatrices of $1-t_0$ and $1-\tau_0$, namely
\beq
\lambda_L^\pm=1-\frac{1}{2}\left[
(\alpha_1+\alpha_4)\pm\sqrt{(\alpha_1-\alpha_4)^2+4\beta^2}\right], \ \ 
\lambda_M^\pm=1-\frac{1}{2}\left[
s(\alpha_2+\alpha_3)\pm\sqrt{s^2(\alpha_2-\alpha_3)^2+4\beta^2}\right]
\label{eigenvals}
\eeq
with the parameters
$\alpha_1+\alpha_4=2gI(|z|^2+m^2)$, $\alpha_1-\alpha_4=2gmI(z^*+z)$ and
$\alpha_2+\alpha_3=2gI(|z|^2-m^2)$, $\alpha_2-\alpha_3=2gmI(z^*-z)$.
For $s=0$ the eigenvalues $\lambda_M^\pm$ are always massive:
\beq
\lambda_M^\pm=1\mp\beta=\cases{
gI(|z|^2+m^2) \cr
2-gI(|z|^2+m^2) \cr
}
\eeq
and for $s=\pm 1$ we consider two special cases, the behavior at the Dirac node and the gapless case:

{\it (I) At the Dirac node $\mu=0$}: 
For the Hamiltonian $H_m$ of Eq. (\ref{gap_ham}) we get the parameter $\beta\sim 1-g(\eta^2+m^2)I$,
and the eigenvalues of Eq. (\ref{eigenvals}) now read
\beq
\lambda_L^\pm=1- gI(\eta^2+m^2)]\mp [1-gI(\eta^2+m^2)] , \ \ 
\lambda_M^\pm=1-s(\eta^2-m^2)gI\pm \sqrt{[1-gI(\eta-m)^2][1-gI(\eta+m)^2]}
\ .
\label{eigenv1}
\eeq
Thus, the ladder contribution $\lambda_L^+$ is always massless, in contrast to $\lambda_L^-$ and the maximally crossed 
contributions, which are all massive. 

{\it (II) Gapless spectrum $m=0$}:
For the Hamiltonian $H_0$ we have $\beta\sim 1-g|z|^2I$, and the eigenvalues read
\beq
\lambda_L^\pm=1-g|z|^2I\mp(1-g|z|^2I) , \ \  \lambda_M^\pm=1-sg|z|^2I\pm (1-g|z|^2I)
\  ,
\label{eigenv2}
\eeq
such that there is a massless mode $\lambda_L^+=0$ for any $s$, like for the Dirac node, and an additional massless mode 
from the maximally crossed contribution $\lambda_M^-=0$ for $s=1$. These results are summarized in 
Table \ref{m_eigenvalues}.

\subsection{Diffusion propagator}

The findings of the previous Section can be used to evaluate the correlation function as
\beq
\langle G_{rr',ab}G^*_{rr',cd}\rangle \approx 
\frac{1}{g_{bd}}({\bf 1}-t)^{-1}_{rr';ac,bd}+\frac{1}{g_{bd}}({\bf 1}-\tau)^{-1}_{rr';ad,bc}
-\frac{1}{g_{bd}}(t^2_{rr';ac,bd} +t_{rr';ac,bd}+2\delta_{rr'}\delta_{ac}\delta_{bd})
\ .
\label{2pgf2}
\eeq
For the long-range behavior with $|r-r'|\sim\infty$ it is sufficient to consider the two-particle 
propagators $({\bf 1}-t_q)^{-1}$ and $({\bf 1}-\tau_q)^{-1}$ for $q\sim0$. Then we can focus on the massless (diffusion) 
modes as the most important contributions to get
\beq
\sim \frac{\eta}{\delta + D_{t,\tau} q^2} \to \frac{\eta}{-i\omega + D_{t,\tau} q^2}
\ ,
\label{diff_prop}
\eeq
where the second expression is obtained by the analytic continuation $\delta\to -i\omega$.
The diffusion coefficients $D_{t,\tau}$ are obtained from the $q$ expansion of the eigenvalues of (\ref{matrices1}) as 
\[
D_t=\frac{g\eta}{2}\sum_r r_\mu^2 [G_{0;r0,11}G^*_{0;0r,11}+G_{0;r0,12}G^*_{0;0r,12}]+o(g^2)
\]
and additionally for $s=1$
\[
D_\tau=\frac{g\eta}{2}\sum_r r_\mu^2 [G_{0;r0,11}G^*_{0;0r,22}+G_{0;r0,12}G^*_{0;0r,12}]+o(g^2)
\ .
\]
Both coefficients agree, at least up to terms of $o(g^2)$, and give us 
\beq
D_t=D_\tau=\frac{g}{4\pi \eta}
\left(1+\frac{1+\zeta^2}{\zeta}\arctan\zeta\right)+o(g^2)
 \ \ \ (\zeta=\mu/\eta)
\ .
\label{diff_const}
\eeq

\begin{table}
\begin{center}
\begin{tabular}{|c|c|c|c|}
\hline  & independent random diagonal elements & random scalar potential & random gap  \\ 
\hline  &  $s=0$  &  $s=1$  &  $s=-1$  \\ 
\hline $\mu=0$ & $\lambda_L^+=0$ &  $\lambda_L^+=0$  & $\lambda_L^+=0$ \\ 
\hline  $m=0$ & $\lambda_L^+=0$  & $\lambda_L^+=0$, $\lambda_M^-=0$ &  $\lambda_L^+=0$ \\ 
\hline  $m,\mu\ne 0$ & none  & none & none \\ 
\hline 
\end{tabular} 
\caption{Vanishing eigenvalues of ${\bf 1}-t_0$ and ${\bf 1}-\tau_0$ (cf. Eqs. (\ref{2pgf_mb}), (\ref{eigenv1}), (\ref{eigenv2})).}
\label{m_eigenvalues}
\end{center}
\end{table}

\section{Corrections to the Boltzmann-Drude conductivity}
\label{sect:correction}

\begin{figure}[t]
\begin{center}
\psfrag{mu^2}{$\zeta^2$}
\psfrag{conductivity}{conductivity $[e^2/\pi h]$}
\includegraphics[width=10cm]{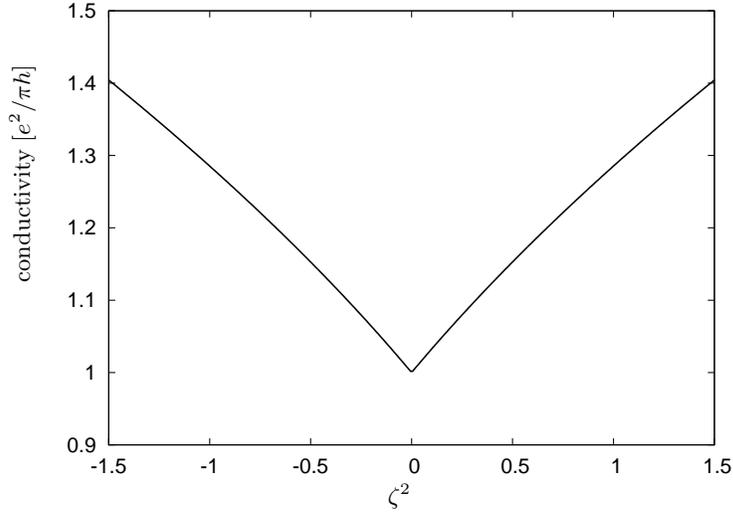} 
\caption{Conductivity as a function of $\zeta^2$ from the expression in Eq. (\ref{ddc}).}
\label{fig:conductivity}
\end{center}
\end{figure}

Next, the results of the WLA will be used to evaluate the quantum corrections to the Boltzmann-Drude conductivity.
The latter is usually calculated in terms of the current-current correlation function for the real part of the 
conductivity \cite{andoetal}
\beq
\sigma_{\mu\mu}\sim \frac{1}{\pi\hbar} \langle
{\rm Tr}\left(j_\mu Gj_\mu G^\dagger\right)\rangle
\ .
\eeq
The advantage of using this expression is that the current-current correlation function 
$\langle{\rm Tr}(\langle j_\mu G j_\mu G^\dagger\rangle)\rangle$ is closely related to the action of the corresponding
nonlinear sigma model $(1/2t)\int {\rm Tr}(\partial_\mu Q\partial_\mu Q)$ \cite{hikami80}, since the current operator $j_\mu$
of a conventional one-band model is proportional to the momentum operator $-i\partial_\mu$. Therefore, the renormalization
of the parameter $t$ is similar to the renormalization of the current-current correlation function. This relation, however, breaks down
for Dirac fermions, where $j_\mu$ is proportional to the Pauli matrix $\sigma_\mu$. For this reason it is not obvious that the
renormalization of the nonlinear sigma model is linked to the renormalization of the current-current correlation function.
In this case it is better to use an alternative 
Kubo formula, which is based on the density-density correlation function \cite{wegner79,ziegler08a}:
\beq
\sigma_{\mu\mu}=-\frac{e^2}{2h}\omega^2\sum_r r_\mu^2Tr_2\langle G_{r0}G^\dagger_{0r}\rangle
\ .
\label{conductivity}
\eeq
This expression is closely related to diffusion and the Einstein relation \cite{ziegler12b}. 
The classical approximation assumes a weak correlation between the two Green's functions $G$, $G^\dagger$ such that we can
factorize the expectation value as $\langle G_{r0}G^\dagger_{0r}\rangle\approx \langle G_{r0}\rangle\langle G^\dagger_{0r}\rangle$
and obtain the Boltzmann-Drude conductivity as
\beq
{\bar\sigma}=-\frac{e^2}{2h}\omega^2\sum_r r_\mu^2Tr_2\langle G_{r0}\rangle\langle G^\dagger_{0r}\rangle
\ .
\eeq
Furthermore, the average one-particles Green's functions are evaluated within the SCBA as 
$\langle G_{r0}\rangle\approx{\bar G}_{r0}$. For the gapless case $m=0$ and for the parameter $\chi=2\mu/\omega$ we can write
\beq
\sum_r r_\mu^2Tr_2\langle G_{r0}\rangle\langle G^\dagger_{0r}\rangle\approx
\sum_r r_\mu^2Tr_2{\bar G}_{r0}{\bar G}^\dagger_{0r}\approx
\cases{
-\frac{1}{\pi\omega^2}\left[1+\frac{1}{4\chi}(1-\chi^2)\log\left(\frac{(1+\chi)^2}{(1-\chi)^2}\right)\right] & for $\omega\gg\eta$ \cr
\frac{1}{4\pi\eta^2}\left[1+\frac{1}{\zeta}(1+\zeta^2)\arctan\zeta
\right] & for $\omega\ll\eta$ \cr
}
\ .
\label{2pgf3}
\eeq
In particular, for $\omega\gg\eta$ the Boltzmann-Drude conductivity reads
\beq
{\bar\sigma}
=\frac{e^2}{2\pi h}\left[1+\frac{1}{4\chi}(1-\chi^2)\log\left(\frac{(1+\chi)^2}{(1-\chi)^2}\right)\right]
\ .
\eeq
This expression is obviously not the Boltzmann-Drude conductivity of a conventional metal with one-band Hamiltonian.
At the Dirac node $\mu=0$ it has a frequency independent conductivity ${\bar\sigma}=e^2/h\pi$  and 
decreases monotonically from $e^2/h \pi$ to zero as we move the Fermi energy $\mu$ away from the Dirac node.
This indicates a cross-over from the optical conductivity of the two-band model at the Dirac node to the 
Boltzmann-Drude behavior of a conventional metal, where the optical conductivity is much lower than in the two-band case.
Here it should be noticed that there are additional corrections for $\omega\gg\eta$, which increase the optical
conductivity to $\pi e^2/8$. These are not taken into account here, since they disappear in the DC limit \cite{ziegler08a}.
For $\omega\ll\eta$, on the other hand, the Boltzmann approximation is invalid, which is reflected by a negative
Boltzmann-Drude conductivity that vanishes like ${\bar\sigma}\propto \omega^2$ .
For this regime we must consider the corrections $\delta \sigma$ which can be evaluated from Eq. (\ref{2pgf2}) as
\beq
\lim_{\omega\to0}\omega^2\sum_{r}(r_\mu-r_\mu')^2\langle G_{rr',ab}G^*_{rr',ab}\rangle \approx 
\frac{1}{g}\lim_{\omega\to0}\omega^2\sum_{r}(r_\mu-r_\mu')^2
[({\bf 1}-t)^{-1}_{rr';aa,bb}+({\bf 1}-\tau)^{-1}_{rr';ab,ba}]
\ .
\label{2pgf2a}
\eeq
With the help of the propagator in Eq. (\ref{diff_prop}) and the diffusion coefficient (\ref{diff_const}) 
the conductivity corrections then read
\beq
\delta\sigma = \frac{e^2}{h}\omega^2\frac{\partial^2}{\partial q_l^2}\frac{\eta/g}{-i\omega +D q^2}\Big|_{q=0}
=\frac{e^2}{h g}2\eta D =\frac{e^2}{2\pi h}\left(1+\frac{1+\zeta^2}{\zeta}\arctan\zeta\right)+o(g^2)
\sim \frac{e^2}{\pi h}(1+\zeta^2/3)
\ .
\label{ddc}
\eeq
There is an additional factor 2 for $s=1$ due to the extra massless mode from $\lambda_M^-=0$ in that case. 
This result, which is depicted in Fig. \ref{fig:conductivity}, agrees with previous calculations based on the WSA 
\cite{ziegler12b} as well as with the V-shape conductivity
with respect to $\mu^2$ in graphene \cite{novoselov05,zhang05}.

\section{Discussion}
\label{sect:discussion}

Our main result is that the ladder and the maximally crossed contributions are quite different for the spinor Hamiltonians, 
in contrast to their agreement for the scalar Hamiltonian of Sect. \ref{sect:scalar}. This situation is reminiscent of the 
Bose-Fermi functional representation of the average two-particle Green's function in the case of Dirac fermions 
with a random gap, where it was observed that the bosonic and the fermionic propagators are distinct \cite{ziegler97}. 
Similar to the expressions in Eq. (\ref{2pgf_mb}), the inverse bosonic two-particle propagator is also a $4\times4$ matrix 
for $a,...,d=1,2$
\[
\delta_{ac}\delta_{bd}-g\int_k G_{0;k,ac}(z)G_{0;k-q,bd}(z)
\ ,
\]
while its fermionic counterpart reads
\[
\delta_{ac}\delta_{bd}-g\int_k G_{0;k,ac}(z)G_{0;-k-q,db}(z)
\ .
\] 
A straightforward calculation shows that only the fermionic propagator has a massless mode, very similar to ${\bf 1}-t_q$ with $s=-1$ of the WLA. 
Therefore, the comparison of the WSA with the WLA sheds some light on the role of Bose-Fermi (or super-) symmetry breaking, 
which can be associated with the difference between ladder diagrams and maximally crossed diagrams in the WLA.
In particular, the massless mode of the WLA agree with the massless fermion mode of the WSA, where the latter is caused
by a spontaneous breaking of a non-Abelian chiral symmetry \cite{ziegler97,ziegler98,ziegler12b}.

It remains to discuss whether or not the structure of the diffusion poles is robust with respect to approximations. 
In the one-band projection of the one-particle Green's function the pole of the second band has been ignored 
\cite{khveshchenko06,tkachov11,schmeltzer13}, since it is energetically too far away from the Fermi surface. For $\mu>0$ we have
\beq
G_{0;k,ab}=U_{1a}^*\frac{1}{\epsilon(k)-z}U_{1b}+U_{2a}^*\frac{1}{-\epsilon(k)-z}U_{2b}
\to U_{1a}^*\frac{1}{\epsilon(k)-z}U_{1b} \ \ \ {\rm with}\ \ U=\frac{1}{\sqrt{2}}\pmatrix{
\kappa^* & 1 \cr
-\kappa^* & 1 \cr
}
\eeq
and $\kappa=(k_1-ik_2)/k\equiv e^{i\Phi(k)}$, which provides the expression
\beq
G_0\approx \frac{1}{\epsilon_k-\mu+i\delta+i\eta}\frac{1}{2}(\sigma_0-\sigma_1k_1/k-\sigma_2 k_2/k)
\ .
\eeq
For the projected Green's function the matrices (\ref{2pgf_mb}) then read
\beq
t_0=\frac{gI}{4}\pmatrix{
1 & 0 & 0 & 1 \cr
0 & s & 0 & 0 \cr
0 & 0 & s & 0 \cr
1 & 0 & 0 & 1 \cr
}, \ \ \ 
\tau_0=
\frac{gI}{4}\pmatrix{
1 & 0 & 0 & 0 \cr
0 & s & 1 & 0 \cr
0 & 1 & s & 0 \cr
0 & 0 & 0 & 1 \cr
}
, \ \ \ I=\int_k\frac{1}{|k^2-z|^2}
\ ,
\eeq
where we get $gI=2-2\delta/\eta$ from the SCBA. Thus the eigenvalues of ${\bf 1}-t_0$ are $0,1,1-s/2,1-s/2$ and 
the eigenvalues of ${\bf 1}-\tau_0$ are 
$1-(s+1)/2, 1-(s-1)/2,1/2,1/2$. Like in the case of the two-band Green's function of Sect. \ref{sect:spinor_gf}
there is one massless mode for ${\bf 1}-t_0$ for any value of $s$ and one massless mode for ${\bf 1}-\tau_0$ if $s=1$.
This indicates that the result of the one-band projection preserves the structure of two-band result of Table 
\ref{m_eigenvalues} in terms of the number of massless modes. The agreement of the results from the one-band projected 
Green's function and the two-band Green's function reflects the fact that the type of diffusive modes is not sensitive 
to the scattering to the second band.

\subsection{Inter-node scattering}

Finally, we briefly discuss the effect of inter-node scattering for the Hamiltonian $H_0$. 
For this purpose we introduce the extended Hamiltonian
\beq
{\bar H}_0=\pmatrix{
H_0 & v\sigma_0 \cr
v\sigma_0 & H_0^* \cr
}
\ ,
\eeq
which describes inter-node scattering by the random scattering terms $v\sigma_0$.

Now we could evaluate the inverse two-particle Green's functions within the WLA and study the vanishing
eigenvalues, which would be associated with diffusive behavior. Alternatively, we can also start from the symmetry
argument and analyze the underlying chiral symmetry whose spontaneous breaking would create a massless mode,
analogously to the treatment of random scattering terms in Ref. \cite{ziegler12b}. Following this concept, 
we first realize that the Hamiltonian ${\bar H}_0$ changes its sign under the transformation in Eq. (\ref{ham0})
\beq
{\bar H}_0\to S{\bar H}_0S=-{\bar H}_0 , \ \ \ S=\pmatrix{
\sigma_3 & 0 \cr
0 & -\sigma_3 \cr
}
\ .
\eeq
According to the general procedure of Ref. \cite{ziegler12b}, this leads for the Hamiltonian 
${\hat H}_0=diag({\bar H}_0,{\bar H}_0)$ to the non-Abelian chiral symmetry
\beq
e^{\hat S}{\hat H}_0e^{\hat S}={\hat H}_0 , \ \ \ {\hat S}=\pmatrix{
0 & \varphi S \cr
\varphi' S & 0 \cr
}
\eeq
for independent continuous parameters $\varphi$, $\varphi'$, since ${\hat H}_0$ and ${\hat S}$ anticommute.
This symmetry is spontaneously broken due to the scattering rate $\eta$, causing the appearance of a massless 
mode. Here it should be noticed though that the appearance of a massless mode is only a necessary but not a sufficient 
condition for a diffusive behavior because the interaction of the nonlinear symmetry fields generated by $\varphi$, 
$\varphi'$, could lead to Anderson localization \cite{hill13b}.

\section{Conclusions}

The calculation of the ladder and maximally crossed contributions of the average two-particle Green's function
has revealed a characteristic diffusion pole structure. Depending on the type of randomness, the ladder contributions
always have one diffusion pole, provided the Fermi energy is at the Dirac node or away from the Dirac node but in the 
absence of a gap. Moreover, for a random scalar potential there is an additional diffusion pole from the maximally 
crossed contributions. All these results require the existence of a nonzero scattering rate, obtained as a solution
of the self-consistent Born relation (SCBA). No diffusion pole have been found away from the Dirac node in the 
presence of a one-particle gap.

The diffusion pole structure of the WLA is identical to that of the WSA, at least for the case of a random gap. 
This enabled us to identify the origin of the diffusion poles with massless modes which are created by spontaneously 
broken symmetries. These are chiral symmetries, associated with the symmetry of the two bands. Such a symmetry exists 
also in the presence of inter-node scattering. Further work is necessary though, to compare the relation between the 
WLA and the WSA for other types of disorder scattering.

The diffusion pole structure of the two-particle Green's function is preserved when we employ a one-band projection 
of the one-particle Green's function by removing one pole of the latter. Although this projection changes the
form of the diffusion coefficient (cf. \cite{khveshchenko06,tkachov11}), it may serve as a good approximation that
reduces the computational effort significantly. 

As already mentioned in the Introduction, the existence of a diffusion pole is a necessary condition but does not 
guarantee a metallic behavior. Higher order terms beyond the ladder and maximally crossed contribution can destroy
diffusion and eventually lead to Anderson localization. This was studied recently in terms of a strong scattering 
expansion \cite{hill13b}, which revealed an exponential decay when the scattering rate $\eta$ is larger than the
band width $E_b$.   

\vskip1cm

\no
Acknowledgement

\no
We are grateful to E. Hankiewicz, D. Schmeltzer and G. Tkachov for inspiring discussions.

\end{document}